\documentclass{aa}
\usepackage{graphicx}
\usepackage{natbib}
\bibpunct{(}{)}{;}{a}{}{,} 

\begin{document}

\title{Microlensing of Strongly Interacting Binary Systems}

\author{V. Bozza\inst{1,3},
        L. Mancini\inst{1,2,3}}
\offprints{V. Bozza, \\
    \email{valboz@sa.infn.it}}
\institute{Dipartimento di Fisica ``E.R. Caianiello'',
           Universit\`{a} di Salerno, I-84081 Baronissi (SA), Italy
           \and Institut f\"{u}r Theoretische Physik der
           Universit\"{a}t Z\"{u}rich, CH-8057 Z\"{u}rich, Switzerland
           \and Istituto Nazionale di Fisica Nucleare, sez. Napoli, Italy}
\date{Received  / Accepted}

\abstract{

Many binary systems are in a state of strong interaction, where
mass exchanges, accretion disks, common envelopes provide
circumstellar matter which can have significant effects in
microlensing light curves of background sources. Such chromatic
absorption effects provide very interesting information on the
nature of this diffuse matter and the physics of the interaction
between the two stars.

\keywords{Gravitational lensing -- Binaries: close --
Circumstellar matter}}

\maketitle

\section{Introduction}

It is well known that about one half of the stars populating our
Galaxy are part of gravitationally bound multiple systems. On the
other hand, binary systems acting as microlenses produce drastic
modifications into microlensing light curves (Mao \& Paczynski
1991). At least $10\%$ of all microlensing events clearly show the
typical signatures of binary lenses, including the presence of
sharp caustic crossing peaks.

A conspicuous fraction of binary systems is in a state of tight
interaction, where several physical phenomena show up and
accompany the basic gravitational rotation. Tidal interactions,
mass exchanges, the appearance of accretion disks and
circumstellar matter around compact objects in their various
shapes widely enrich the phenomenology of these systems. For a
detailed review on strongly interacting binary systems (hereafter
SIBS), we refer the reader to (Carroll \& Ostlie 1996), where a
full chapter is devoted to the subject.

In the present paper, we want to point out at all the possible
signatures of SIBS in microlensing observations, with particular
reference to chromatic absorption by circumstellar matter in
microlensing light curves. These effects can be appreciated in
careful observations and could provide an alternative tool to
detect SIBS and determine their characteristics in a qualitative
and quantitative way.

In the binary lens equation, two scales are present: the total
Einstein radius $R_\mathrm{E}$, proportional to the square root of
the total mass, and the separation between the binaries. In
practice, the system will best show up its binary nature through
microlensing when these two scales are of the same order. Of
course, not all close binary systems will reveal their nature in a
microlensing event. The separation between the two stars cannot be
lower than $0.1 R_\mathrm{E}$ (Gaudi \& Gould 1997). If the two
stars are too close, they will effectively behave as a single lens
and their binary nature would be hidden to microlensing. Typical
Einstein radii in galactic observations are of the order of AU,
which are compatible with most classes of interacting binary
systems, characterized by separations of the order of tenths of
AU. However, if we also take into account astrometric observations
of the center of light deformations during microlensing, the
efficiency in the detection of close binaries may be decisively
increased, so that even binaries with separations of the order
$0.01 R_\mathrm{E}$ can be detected (Chang \& Han 1999).

The rotation periods of interacting binaries may range from few
hours to several months. As regards those binaries which can yield
distinctive features of binary microlensing (namely those which
have separations larger than 0.1, 0.2 AU), for solar mass stars,
the typical periods are of the order of some tens of days. This
means that the rotation of the lens should be often included in a
detailed study of the microlensing event. A comprehensive study of
this problem has been already performed by Dominik (1998) and will
not be discussed here.

\section{Diffuse matter in binary systems and microlensing}

If the size of one of the stars in the binary system exceeds its
Roche lobe, a mass transfer process towards the other star
initiates. If this happens, the star losing mass through its Roche
lobe will be referred to as {\it secondary} star, while the star
accreting mass will be the {\it primary} star. The matter leaving
the secondary star spirals around the primary forming an {\it
accretion disk}, whose size depends on the geometrical and
dynamical properties of the system. The comprehension of the
accretion disk around a compact object is important to obtain
precious information about the nature of the collapsed object and
its physical properties.

The matter in the accretion disk is hydrogen supplied by the
external layers of the secondary star's atmosphere. In typical
situations, the temperature of the accretion disk allows the
presence of neutral hydrogen. In degenerate primaries (neutron
stars or black holes), however, the temperature of the inner
regions of the accretion disk may rise to $10^6$ K, favouring a
full ionization of matter.

The secondary star may also spontaneously expel mass as a natural
stage in its evolutionary life or as a consequence of pulsations
induced by tidal interactions.  The mass lost by the the secondary
star may set between the two stars, forming a circumstellar
envelope of neutral hydrogen and dust (Dumm et al. 1999). If the
ejection velocity is high enough, then a common envelope surrounds
both stars for long periods of the life of the system.

From this discussion, we learn that the presence of gas and dust
in binary systems is a typical product of strong gravitational
interactions. Generically, in microlensing observations, the
presence of diffuse matter shows up through chromatic absorption
effects.

\subsection{Chromatic Absorption}

The light passing through the gas and the dust within the binary
system is attenuated by scattering and absorption phenomena. The
Lambert's law
\begin{equation}
\label{Lambert} I_{\lambda}=I_{\lambda,0} e^{-\tau_{\lambda}},
\end{equation}
describes the suppression experienced by light travelling through
gas. $I_{\lambda,0}$ is the original intensity of light,
$I_{\lambda}$ is the transmitted light and $\tau_{\lambda}$ is the
optical depth of the cloud, given by
\begin{equation}
\tau_\lambda=\sum\limits_n k_{\lambda,n}\sigma_n (\mathbf{x}),
\end{equation}
where $k_{\lambda,n}$ is the absorption coefficient and $\sigma_n
(\mathbf{x})$ is the column density of the gas component labeled
by $n$ at projected position $\mathbf{x}$.

Summing up, each image will have a total amplification given by
the geometrical magnification due to gravitational light bending,
weighted by the transmission coefficient $e^{-\tau_\lambda}$.

The absorption coefficient can be expressed as the cross section
over the mass of the elementary gas/dust particle. For instance,
H$_2$ Rayleigh scattering cross section is $2.24\times 10^{-27}$
cm$^2$ at $\lambda=4400$ \AA. To appreciate a $50\%$ absorption,
the column density should be about $10^4$ Kg/m$^2$, which is the
column density of one thousandth solar masses distributed on a
uniform disk with radius equal to 2.7 AU. However, slight changes
of the column density or the chemical nature of the gas may drive
the absorption towards complete occultation or complete
transparency. What is interesting for us is that the typical
densities of circumstellar matter in interacting binary systems
are compatible with partial or total absorption of background
light and thus produce observable effects.

\begin{figure*}
 \resizebox{\hsize}{!}{\includegraphics{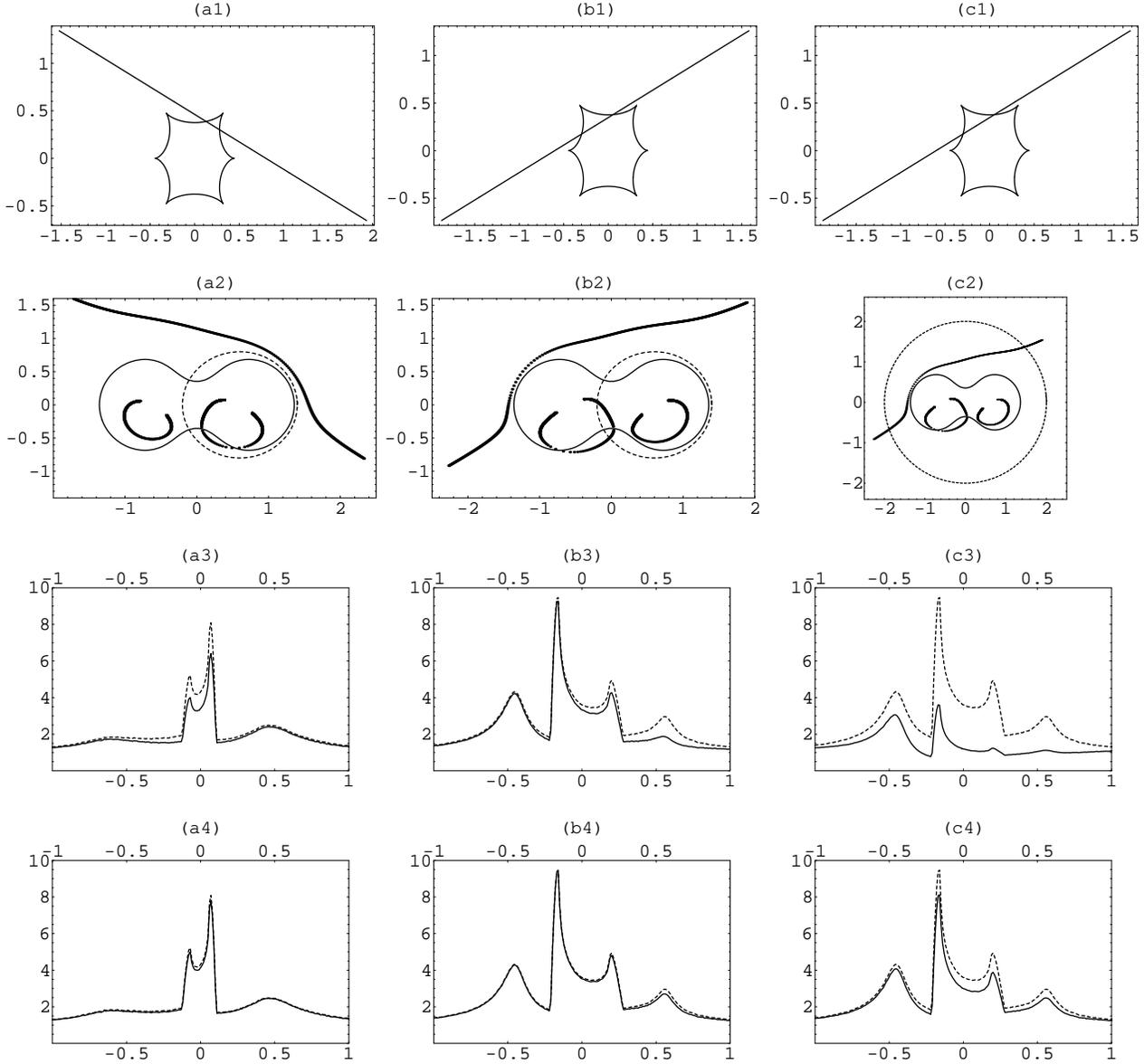}}
 \caption{Three examples of microlensing light curves for the lens described
  in the text. Each column describes a different case.
  In the top row there are the source trajectories and the caustic.
  In the second row there are the images, the critical curve and the border of the disk
  of diffuse matter (dashed). All distances are measured in
  Einstein radii.
  In the third row there are the three amplification curves in blue band ($\lambda=0.44
  \mathrm{\mu m}$), along with the light curves without absorption (dashed).
  On the vertical axis there is the amplification; on the horizontal there is
  time in Einstein units.
  In the bottom row there are the amplification curves in red band ($\lambda=0.7 \mathrm{\mu m}$).}
 \label{Fig Lig}
\end{figure*}

In binary microlensing, three images are normally present, while
two more are created when the source is inside a caustic. Only the
images which are formed inside the diffuse matter envelope are
affected by absorption, while the others remain undisturbed. Of
course, a complete classification of all possible light curve is
beyond the purpose of this letter, since accretion disks and
gaseous envelopes may appear in an extraordinary variety of shapes
and can be oriented in whatever way with respect to the observer.
However, in Fig. \ref{Fig Lig}, we analyze some simple situations
to give an idea of the anomalies that could appear in binary
microlensing light curves, due to the presence of diffuse matter.

We consider a binary system composed by two half-solar mass stars,
separated by 4.3 AU, at a distance of 8 kpc from the observer.
With these values, rotational effects are negligible and will not
be considered, so that we are allowed to concentrate just on
chromatic absorption effects. The lensed source is 10 kpc distant
and has a radius of 15.5 R$_\odot$. One of the lenses (the right
one in the figures) is surrounded by a gaseous disk with projected
density profile
\begin{equation}
\sigma(r)=1.27 M_\mathrm{d} \frac{\left(r_\mathrm{d}^2-r^2
\right)^3}{r_\mathrm{d}^8} \label{sigma}
\end{equation}
where $r_\mathrm{d}=2.87$ AU is the radius of the disk, $r$ is the
distance from the center of the right star, $M_\mathrm{d}=0.01
\mathrm{M}_\odot$ is the total mass of the disk. The choice of
this density profile has been done just for computational reasons.
In fact, the density profiles of circumstellar matter within
interacting binaries are known only in some specific cases. This
prevents us from operating more physical choices. However, since
we are only interested in a qualitative description of the effects
on microlensing light curves, we are allowed to pioneer the
problem by Eq. (\ref{sigma}), also neglecting tidal distortions,
in a first extent. The gas in the disk will be taken to be a
homogeneous mixture of H$_{2}$-He with $24\%$ He by mass, so that
the absorption coefficient is $k_{\lambda}=5.16\times 10^{-5}$
m$^2$/Kg at $\lambda= 4400$ \AA.

The distinctive characteristic of absorption by diffuse matter is
given by its dependence on the wavelength of the light. Rayleigh
scattering decreases as $\lambda^{-4}$, so that all the anomalies
would be harder in blue and softer in red. This is very important
for an unambiguous identification of the effect and we shall plot
light curves both in blue ($\lambda=0.44 \mathrm{\mu m}$) and in
red band ($\lambda=0.7 \mathrm{\mu m}$).

Gravitational and refractive bending induced by gaseous matter are
generally negligible with respect to absorption for our parameters
(Bozza et al. 2002) and will not be considered. We have chosen
three examples of light curves showing the main anomalies that can
be induced by the presence of diffuse matter in the system.

\begin{itemize}
\item[$\bullet$]{
Fig. \ref{Fig Lig} Col. (a) shows a microlensing light curve where
the source crosses a caustic near a cusp. The right secondary
image is destroyed and regenerated in two points which are
completely inside the absorbing disk. This causes the two caustic
crossing peaks to be highly suppressed in blue band, while very
slight modifications appear in red. The secondary non-caustic
peaks stay almost unmodified in both colors. }

\item[$\bullet$]{ In Fig. \ref{Fig Lig} Col. (b) the left secondary image
is regenerated outside the disk and destroyed inside. As a result,
only the second caustic crossing peak is depressed. Moreover, the
right secondary peak, caused by the close approach between the
source and the right top cusp, is suppressed as well. This is
because the right secondary image, which is highly magnified by
gravitational lensing during this peak, is heavily absorbed by the
disk. These depressions almost disappear in red band. It is also
possible to draw curves where no caustic-crossing peaks are
suppressed but one secondary peak is. }

\item[$\bullet$]{ The scenario changes if also the principal image enters the disk.
To show this case, in Figs \ref{Fig Lig} Col. (c) we have enlarged
the disk of diffuse matter up to 7 AU, recentering it on the
center of mass (again we are neglecting any ellipsoidal
distortion). The curve is gradually suppressed in every region
from the outer to the inner portions, as the principal image
reaches the denser inner regions. The red band light curve still
resembles an ordinary binary microlensing curve, but the blue band
one is heavily suppressed.
 }

\end{itemize}

By these three examples we do not mean to give an exhaustive
picture, but at least we can learn that if the diffuse matter
remains in the gravitational influence of one mass only, then we
can expect slight distortions on the original microlensing light
curve, which can be restricted to a limited number of peaks. If
the diffuse matter involves the whole system, then a general
suppression of the microlensing event in all its parts should be
expected.

If the deviations from standard binary microlensing light curves
are small, it should not be difficult to identify these events as
binary microlensing and then examine {\it a posteriori} all
possible anomalies comparing the light curves in different bands.
If the modifications are too strong and, in particular, depress
the event towards the background noise, then it may become very
difficult to identify the event as microlensing. However, (as in
Figs. \ref{Fig Lig} Col. (c)) it may happen that events strongly
absorbed in blue are still recognizable as binary microlensing in
red, so that they can be recovered anyway. Of course, to perform
this kind of searches, it is necessary to relax the selection
procedures based on microlensing achromaticity.

The chromaticity effects of SIBS are the key to identify them.
Chromatic effects in caustic crossing peaks also arise when
limb-darkening of finite size sources is taken into account (Han
\& Park 2001). However, in this case, the color shifts towards
blue, while diffuse matter in binary systems provokes a reddening
of the source light. The two effects are thus clearly
distinguishable.

In exceptional cases, when one of the stars in the binary system
is a low density red giant, occultations of one secondary image
may occur (Bromley 1996; Agol 2002). The effects on the light
curve are similar to those shown in Fig. \ref{Fig Lig} Cols.
(a)-(b) for diffuse matter. The only difference is in the density
profile that should be very close to a step function.

\section{Discussion}

The presence of diffuse matter within SIBS shows up through
absorption of one or more microlensing images. If the diffuse
matter is localized around one star, we expect a partial
depression of caustic-crossing or secondary peaks, depending on
the geometry of the event. If the matter englobes the whole
system, then all microlensing images are simultaneously affected
and a general suppression of the curve occurs, which may
significantly alter the original microlensing curve. Since the
modifications are more evident in blue than in red, a careful
analysis in different color bands is mandatory to unambiguously
detect and study these events in a serious way. Moreover, if it is
possible to take a spectrum during the microlensing event, the
appearance of characteristic absorption lines will be the clear
signature of the presence of the diffuse medium. In this case, we
would also have a way to analyze the chemical nature of this
diffuse matter.

Microlensing may provide a very powerful tool for a deep physical
investigation of SIBS. It may help to estimate the distribution of
diffuse matter in these systems, also retrieving information on
their density profiles. Eventual turbulent density fluctuations
would be apparent as random noise on the light curve if their
length scales are comparable to the extension of the absorbed
images. In this case, a statistical study of this noise would
provide information on these fluctuations. Moreover, microlensing
comes with its standard information on the mass ratio and the
separation of the stars along with other parameters which can be
useful to give a complete characterization of the system.

An important observation is that interacting binaries are detected
by traditional optical observations only when their orbital plane
is edge-on with respect to the observer (unless they are in the
solar neighbourhood, so that the rotation can be detected by
astrometric measurements, or the two components are separated
interferometrically). The binary nature of the object is then
revealed by the mutual eclipses produced by rotation, or by
spectroscopical measurements of radial velocities. If the orbital
plane is disposed face-on with respect to the observer, it is
necessary to resort to indirect proofs (e.g. X-ray emission from
accretion disk, cataclysmic variables). In microlensing surveys,
instead, both situations are practically detectable with the same
chances. It is thus possible to enlarge the sample of known binary
systems and study them from a different perspective.

\begin{acknowledgements}

We wish to thank Philippe Jetzer and Gaetano Scarpetta for helpful
comments.

\end{acknowledgements}


\bibliographystyle{aa}

\end{document}